\documentclass[aps,prd,preprint,superscriptaddress,tightenlines,nofootinbib]{revtex4}
\usepackage{graphicx}
\usepackage{dcolumn}
\usepackage{bm}

\newcommand{\mathscr}[1]{{\cal #1}}

\newcommand{\etal}{{\it et al.}}

\begin{document}

\preprint{CLNS 06/1967}       
\preprint{CLEO 06-12}         

\title{A Study of the Decays $D^0 \to \pi^{-} \lowercase{e}^{+}
  \nu_e$, $D^0 \to K^{-} \lowercase{e}^{+} \nu_e$, $D^{+} \to \pi^0
  \lowercase{e}^{+} \nu_e$, and $D^{+} \to \bar{K}^0 \lowercase{e}^{+} \nu_e$}

\author{D.~Cronin-Hennessy}
\author{K.~Y.~Gao}
\author{D.~T.~Gong}
\author{J.~Hietala}
\author{Y.~Kubota}
\author{T.~Klein}
\author{B.~W.~Lang}
\author{R.~Poling}
\author{A.~W.~Scott}
\author{A.~Smith}
\author{P.~Zweber}
\affiliation{University of Minnesota, Minneapolis, Minnesota 55455}
% ...
\collaboration{CLEO Collaboration} 
\noaffiliation

\date{December 5, 2007}

\begin{abstract} 
Using 1.8 million $D\bar{D}$ pairs and a neutrino reconstruction
technique, we have studied the
decays $D^0 \to K^- e^+ \nu_e$, $D^0 \to \pi^- e^+ \nu_e$, $D^{+} \to \bar{K}^0 e^+
\nu_e$, and $D^{+} \to \pi^0 e^+ \nu_e$. We find 
${\cal B}(D^0 \to \pi^- e^+ \nu_e) = 0.299(11)(9)\%$, 
${\cal B}(D^+ \to \pi^0 e^+ \nu_e) = 0.373(22)(13) \%$,
${\cal B}(D^0 \to K^- e^+ \nu_e) = 3.56(3)(9) \%$, and 
${\cal B}(D^+ \to \bar{K}^0 e^+ \nu_e) = 8.53(13)(23) \%$.
In addition, form factors are studied through fits to the partial
branching fractions obtained
in five $q^2$ ranges.  By combining our results with recent unquenched lattice calculations,
we obtain $|V_{cd}| = 0.217(9)(4)(23)$ and $|V_{cs}| = 1.015(10)(11)(106)$.
\end{abstract}

\maketitle

Study of the semileptonic decays of $D$ mesons plays an important role in
our understanding of the CKM matrix~\cite{km_paper}.  These decays allow
determination of $|V_{cs}|$ and $|V_{cd}|$ by combining
measured branching fractions with results from unquenched lattice
QCD (LQCD) calculations~\cite{fnalqcd}. With improved branching
fraction precision these measurements also provide rigorous tests
of LQCD \cite{yellowbook}.  
The tests can be approached by assuming
unitarity of the CKM matrix and comparing the constrained matrix elements
\cite{pdg} to measured elements, or by comparing the measured and
calculated ratios of semileptonic and purely leptonic branching
fractions, which are independent of CKM matrix elements.
Verification of LQCD calculations at the few percent level will 
provide validation for their application to the $B$ system.

This Letter presents a study of the $D^0 \to K^- e^+ \nu_e$, $D^0 \to \pi^-
e^+ \nu_e$, $D^{+} \to \bar{K}^0 e^+\nu_e$, and $D^{+} \to \pi^0 e^+ \nu_e$ 
decay modes (charge conjugate modes implied). A more detailed description
of the analysis is provided in a companion article~\cite{PRD}.
The results are based on 281~pb$^{-1}$ of $e^+e^-$ data at the $\psi (3770)$
resonance ($1.8$ million $D\bar{D}$ pairs)
collected with the CLEO-c detector at the Cornell Electron Storage Ring
(CESR)~\cite{detector}. The sample is a superset of the data used to obtain the first
CLEO-c semileptonic branching fraction measurements~\cite{dsemilep}.    
For each mode we determine the partial branching fractions in
five $q^2=m_{\ell\nu}^2$ ranges. Summing the rates yields the
total branching fraction; fitting the rates constrains
form factor shapes; comparing to LQCD calculations~\cite{fnalqcd}
determines the CKM elements $|V_{cd}|$ and $|V_{cs}|$. 

The analysis technique rests upon association of the missing energy and momentum 
in an event with the neutrino four-momentum \cite{nu_recon_other}, enabled by the hermeticity and excellent resolution of the CLEO-c detector. 
Charged
particles are detected over $93\%$ of the solid angle by two wire
tracking chambers within a $1.0$~T solenoid magnet. The momentum
resolution is $0.6\%$ at $800$ MeV/$c$. 
Specific ionization and a ring imaging \v Cerenkov
detector (RICH) provide particle identification; a CsI(Tl)
electromagnetic calorimeter provides photon detection over $93\%$ of $4\pi$ 
and a $\pi^0$ mass resolution of $\sim$$6$~MeV/$c^2$. 

Electron candidates are identified above $200$ MeV$/c$  over
$90\%$ of the solid angle by combining
information from specific ionization with calorimetric, RICH and
tracking measurements. 
To reduce our sensitivity to final state radiation (FSR), we add
photons within 3.5$^\circ$ of the electron flight direction to the
electron momentum.

Signal charged pions and kaons are identified using specific ionization and RICH
measurements. A $\pi^0$ candidate must have a $\gamma\gamma$ mass within
2.5 standard deviations ($\sigma$) of the $\pi^0$ mass. $K_S^0$ candidates are
reconstructed using a vertex fit to candidate $\pi^+\pi^-$
daughter tracks. The $\pi^+\pi^-$ mass must  be within 4.5$\sigma$ of the $K_S^0$ mass.

The missing four-momentum in an event is given by $p_{\mathrm{miss}} =
(E_{\mathrm{miss}},\vec{p}_{\mathrm{miss}})= 
 p_{\mathrm{total}} - \sum p_{\mathrm{charged}} - \sum
 p_{\mathrm{neutral}}$, where the event
four-momentum $p_{\mathrm{total}}$  is known from the energy
and crossing angle of the CESR beams. The charged
and neutral particles included in the sums pass selection criteria designed
to achieve the best possible $|\vec{p}_{\mathrm{miss}}|$ resolution by
balancing the efficiency for detecting true particles against the
rejection of false ones~\cite{PRD}. 

Association of the missing four-momentum with the neutrino
is only valid if the event contains no more than one
neutrino and all true particles are detected. We thus exclude
events that have either more than one electron or non-zero
net charge. The core $|\vec{p}_{\mathrm{miss}}|$ resolution in signal Monte
Carlo (MC) simulated events satisfying these criteria is $\sim$$15$ MeV$/c$.  

To select signal events we require that the 
$M_{\mathrm{miss}}^2 \equiv E_{\mathrm{miss}}^2 -
|\vec{p}_{\mathrm{miss}}|^2$ be consistent with 
a massless neutrino. Using MC simulations that included all resolution
effects, candidate selection was found to be optimized by requiring 
$\left|M_{\mathrm{miss}}^2 \slash 2|\vec{p}_{\mathrm{miss}}|\right| < 0.2$ GeV/$c^3$.  
Since $|\vec{p}_{\mathrm{miss}}|$ resolution is roughly half 
that of $E_{\mathrm{miss}}$, in subsequent calculations we take 
$p_{\nu} \equiv (|\vec{p}_{\mathrm{miss}}|,\vec{p}_{\mathrm{miss}})$.

Semileptonic decays $D \to h e \nu_e$, where $h = \pi$ or $K$, are
identified using four-momentum conservation.
Specifically, reconstructed $D$ candidates are selected based
on $\Delta E \equiv (E_h + E_e + E_\nu) - E_{\mathrm{beam}}$ (expected to
be close to zero within our $20$ MeV resolution). 
The $D$ momentum constraint is recast as the beam-constrained mass $M_{\mathrm{bc}}$, defined below, which peaks near the $D$ mass for signal.

Selection criteria were optimized with independent MC samples.
Background sources include events with hadrons misidentified as
electrons (fake electrons), non-charm continuum production, 
and $D\bar{D}$ processes other than signal. 
The optimal energy requirement was determined to be $-0.06$ $< \Delta
E < 0.10$ GeV. For the Cabibbo-favored modes, the background remaining
after this selection is only a few percent of the signal.  
For the Cabibbo-suppressed modes, significant background remains from
signal-mode cross-feed and from the related modes
$D^+ \to K_L^0 e^+ \nu_e$ and $D^+ \to K_S^0(\pi^0\pi^0) e^+ \nu_e$, where 
the $(\pi^0\pi^0)$ indicates the $K_S^0$ decay mode. Restricting the $\Delta E$ of the non-signal side of
the event reduces these backgrounds. We also require $D^+ \to \pi^0 e^+ \nu_e$
candidates to have the smallest $|\Delta E|$ of any
final state candidate in the event, and that these events contain no 
reconstructed $D^0 \to K^- e^+ \nu_e$ candidate. The average background 
fraction ($q^2$--dependent) in the pion modes is about 20\%.

Since the $|\vec{p}_\nu|$ resolution dominates $\Delta E$ resolution, we improve
our $p_\nu$ measurement by scaling it 
by the factor $\zeta$ satisfying $\Delta E = (E_h + E_e + \zeta E_\nu) - E_{\mathrm{beam}}=0$.
We use $\zeta \vec{p}_\nu$ to calculate 
$M_{\mathrm{bc}} \equiv \sqrt{E_{\mathrm{beam}}^2 - |\vec{p}_{h} + \vec{p}_{e} + \zeta\vec{p}_\nu|^2}$
with a resolution of 4 MeV$/c^2$.
We calculate $q^2 \equiv (p_{\nu} + p_e)^2$ with a
resolution of 0.01 GeV$^2/c^4$, independent of
$q^2$. 

To extract branching fraction information we perform a simultaneous maximum likelihood 
fit~\cite{bbfit} to the $M_{\mathrm{bc}}$ 
distributions of the four signal modes in five $q^2$ ranges.
The simultaneity automatically
provides self-consistent rates for misreconstruction of one signal process
as another (cross-feed) and for background from the related $K^0$ processes.
The $M_{\mathrm{bc}}$ 
distribution is divided into 14 uniform bins over the range 
$1.794 < M_{\mathrm{bc}} < 1.878$ GeV$/c^2$.
To simplify the statistical interpretation of our results we limit the
number of multiple entries per event:  
a given event can contribute to at most one $D^0$ and one $D^+$ final state.
For multiple $D^0$ or $D^+$ candidates with $M_{\mathrm{bc}} >
1.794$ GeV$/c^2$, we choose 
the one with the smallest $|\Delta E|$, independent of $q^2$.

We fit the data to the signal and five background components. The signal mode MC components are based on EvtGen \cite{EvtGen} with modified pole-model (BK parameterization) form factors~\cite{BKparam} and parameters from the most recent
unquenched LQCD calculation~\cite{fnalqcd}. 
Several corrections, relating to inclusive $D$ decay and reconstruction
(see Ref.~\cite{PRD}), are applied to our GEANT-based \cite{GEANT} MC
samples. These lead to few percent (or less) changes in the measured
yields, and are determined precisely enough (using a large $D\bar{D}$ sample
with one fully-reconstructed $D$ meson per event) to yield sub-percent systematic uncertainties.
We are also sensitive to the signal efficiency and kinematic distortions due to FSR. 
Based on the angular and energy distributions for FSR photons, we correct our 
signal MC, generated with PHOTOS~\cite{photos} without interference, to the
leading-order KLOR~\cite{klor} calculations applied to charm decay.

To reduce our sensitivity to form factors,
we extract an independent rate for each of the five $q^2$ intervals
corresponding to the reconstructed $q^2$ ranges in each mode 
(a total of 20 yields). 

\begingroup
\squeezetable
\begin{table*}[t]
\caption{Branching fractions, isospin ratios, and form factor
  parameters (isospin corrected for the $\pi^0 e^+ \nu$ mode). Errors are (stat.)(syst.) or
  (stat.)(syst.)(theor.). Correlation coefficients (from combined statistical and systematic uncertainties) for variables in any
  two (three) preceding columns are given by $\rho$ ($\rho_{ij}$). For the
  $a_i$ we have assumed $|V_{cs}| = 0.976$ and $|V_{cd}| = 0.224$.}
\label{table:bf}
\begin{ruledtabular}
\begin{tabular}{llllllllll}
$q^2$ (GeV$^2/c^4$)                 & $< 0.4$        &  $0.4- 0.8$    & $0.8 - 1.2$   & \multicolumn{3}{l}{$1.2 - 1.6$}     & $ \ge 1.6$    & Total           & $|V_{cq}|$\\
${\cal B}(\pi^- e^+ \nu_e)(\%)$     & $0.070(5)(3)$  &  $0.059(5)(2)$ & $0.060(5)(2)$ & \multicolumn{3}{l}{$0.044(4)(2)$}   & $0.066(5)(2)$ & $0.299(11)(9)$  & 0.218(11)(5)(23)\\
${\cal B}(\pi^0 e^+ \nu_e)(\%)$     & $0.084(10)(4)$ & $0.097(11)(4)$ & $0.062(9)(3)$ & \multicolumn{3}{l}{$0.063(10)(2)$}  & $0.067(11)(3)$& $0.373(22)(13)$ & 0.216(17)(6)(23)\\
${\cal B}(K^- e^+ \nu_e)(\%)$       & $1.441(21)(35)$& $1.048(18)(28)$&$0.681(15)(18)$& \multicolumn{3}{l}{$0.340(11)(10)$} & $0.048(5)(12)$& $3.557(33)(90)$ & 1.023(13)(13)(107) \\
${\cal B}(\bar{K}^0 e^+ \nu_e)(\%)$ & $3.436(82)(93)$& $2.544(73)(69)$&$1.589(58)(44)$& \multicolumn{3}{l}{$0.821(42)(24)$} & $0.139(18)(5)$& $8.53(13)(23)$  & 1.004(20)(15)(105)\\
$I_\pi$                             & $2.12(31)(9)$ & $1.54(22)(7)$  & $2.47(43)(13)$ & \multicolumn{3}{l}{$1.78(32)(7)$}   & $2.48(45)(13)$& $2.03(14)(8)$ & \\
$I_K$                               & $1.06(3)(3)$   & $1.04(4)(3)$   & $1.09(5)(3)$  & \multicolumn{3}{l}{$1.05(6)(4)$}    & $0.88(15)(3)$ & $1.06(2)(3)$ & \\  
\hline
& \multicolumn{9}{c}{Series Parameterization} \\
Decay                 & $a_0$       & $a_1$        & $a_2$        & $\rho_{01}$ & $\rho_{02}$ & $\rho_{12}$ & $|V_{cq}|f_+(0)$ & $1 + 1\slash \beta - \delta$  & $\rho$ \\
$\pi^- e^+ \nu_e$     & 0.044(2)(1) & -0.18(7)(2)  & -0.03(35)(12)& 0.81$\;$    & 0.71$\;$    & 0.96$\;$    & 0.140(7)(3)    & 1.30(37)(12)                    &  -0.85 \\
$\pi^0 e^+ \nu_e$     & 0.044(3)(1) & -0.23(11)(2) & -0.60(57)(15)& 0.80        & 0.67        & 0.95        & 0.138(11)(4)   & 1.58(60)(13)                    & -0.86 \\
$K^- e^+ \nu_e$       & 0.0234(3)(3)& -0.009(21)(7)&  0.52(28)(6) & 0.62        & 0.56        & 0.96        & 0.747(9)(9)   & 0.62(13)(4)                     & -0.62 \\
$\bar{K}^0 e^+ \nu_e$ & 0.0224(4)(3)&  0.009(32)(7)&  0.76(42)(8) & 0.72        & 0.64        & 0.96        & 0.733(14)(11)  & 0.51(20)(4)                     & -0.72 \\
\hline
& \multicolumn{3}{c}{Simple Pole Model} & \multicolumn{5}{c}{Modified
  Pole Model} & \\
Decay                 & $|V_{cq}|f_+(0)$ & $m_{\mathrm{pole}}$ (GeV$/c^2$) & $\rho$ & \multicolumn{3}{l}{$|V_{cq}|f_+(0)$} & $\alpha$    & $\rho$ \\
$\pi^- e^+ \nu_e$     & 0.146(4)(2)      & 1.87(3)(1)                      & 0.63   & \multicolumn{3}{l}{0.142(4)(2)}      & 0.37(8)(3)  & -0.75 & \\
$\pi^0 e^+ \nu_e$     & 0.149(6)(3)      & 1.97(7)(2)                      & 0.65   & \multicolumn{3}{l}{0.147(7)(4)}      & 0.14(16)(4) & -0.75 & \\
$K^- e^+ \nu_e$       & 0.735(5)(9)      & 1.97(3)(1)                      & 0.36   & \multicolumn{3}{l}{0.732(6)(9)}      & 0.21(5)(3)  & -0.42 & \\
$\bar{K}^0 e^+ \nu_e$ & 0.710(8)(10)     & 1.96(4)(2)                      & 0.53   & \multicolumn{3}{l}{0.708(9)(10)}     & 0.22(8)(3)  & -0.59 & \\
\end{tabular}
\end{ruledtabular}
\end{table*}
\endgroup

We also use MC samples to describe the $D\bar{D}$ background and continuum 
contributions. We absolutely scale the continuum components  
according to their cross sections at the $\psi(3770)$ and the measured data luminosity. 
The non-signal $D\bar{D}$ sample was generated using EvtGen, with
decay parameters updated to reflect our best knowledge of $D$ meson decays.  
This component floats separately for each reconstructed final state, but
is fixed over the five $q^2$ regions within that state, thereby
reducing our sensitivity to inaccuracies in the $D$ decay model. 
Finally, we input signal MC components for 
$D^+ \to K_L^0 e^+ \nu_e$ and $D^+ \to K_S^0(\pi^0\pi^0)e^+ \nu_e$, 
whose rates in each $q^2$
region are proportional to those for the reconstructed $D^+ \to K_S^0e^+ \nu_e$ mode with $K_S^0\rightarrow\pi^+\pi^-$.  

The contributions of events with fake electrons are evaluated by
weighting hadron-momentum spectra in candidate events with
misidentification probabilities measured in other CLEO-c data. This component is 
included with a fixed normalization in the fit.   

We allow the fit to adjust the
$M_{\mathrm{bc}}$ resolution in the $D^0 \to \pi^- e^+ \nu_e$, 
$D^0 \to K^- e^+ \nu_e$, and $D^+ \to \pi^0 e^+ \nu_e$ modes  by
applying a Gaussian smear to these distributions.  The result is that the
signal MC $M_{\mathrm{bc}}$ resolution in these modes, $\sim$3.5
MeV$/c^2$, is increased to match the data resolution of $\sim$4 MeV$/c^2$.

\begin{figure}[tb]
\begin{center}
\hspace{0.0cm}
\includegraphics[width=8.5cm]{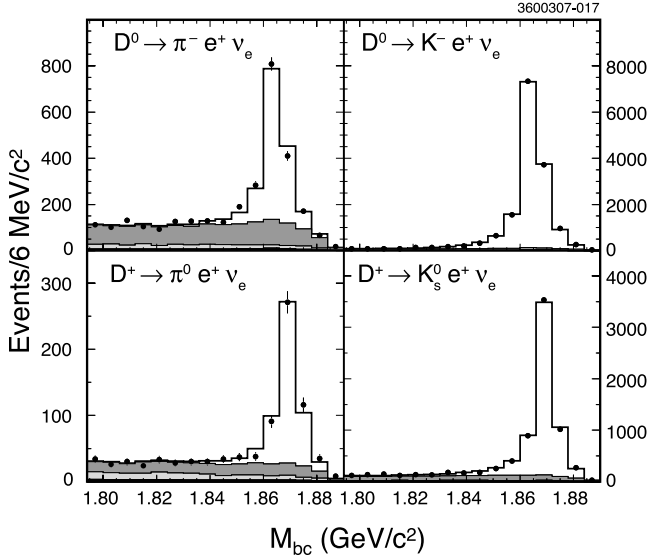} 
\caption{ The reconstructed $M_{\mathrm{bc}}$ distributions, integrated
over $q^2$, for data (points) and components (histograms) from the
fit (see text): signal MC (clear), cross-feed and non-signal $D\bar{D}$ MC
(gray), continuum MC (light gray). The $e^+$ fake component (black) is
negligible on this scale. }
\label{fig:mbc_all}
\end{center}
\end{figure}

The resulting $M_{\mathrm{bc}}$ distributions, integrated over $q^2$, are
shown in Fig.~\ref{fig:mbc_all}
with the fit results (from $D$ and $\bar{D}$ decays) overlaid. 
The value of the likelihood for this fit
is $-2\ln \mathscr{L} = 275.5$ for $280 - 27 = 253$ degrees of freedom.

We obtain branching fractions (see
Table~\ref{table:bf}) for each $q^2$ region 
by
combining the efficiency-corrected yields from the fit with the number
of $D^0\bar{D}^0$ ($N_{D^0\bar{D}^0}$) and $D^+D^-$ ($N_{D^+D^-}$) pairs 
for our sample.  An independent study of 
hadronic $D$ decays~\cite{Dobbs:2007zt} finds 
$N_{D^0\bar{D}^0} = 1.031(16) \times 10^6$
and $N_{D^+D^-} = 0.819(13) \times 10^6$. We
find ratios of branching fractions $R_0 = {\cal B}(D^0 \to \pi^- e^+
\nu_e)\slash{\cal B}(D^0 \to K^- e^+ \nu_e)=8.41(32)(13)\%$ and $R_+ = {\cal B}(D^+ \to
\pi^0 e^+ \nu_e)\slash{\cal B}(D^+ \to \bar{K}^0 e^+ \nu_e)=4.37(27)(12)\%$. 
Table~\ref{table:bf} also lists the partial-width ratios $I_\pi =
\Gamma(D^0 \to \pi^- e^+ \nu_e)\slash \Gamma(D^+ \to \pi^0 e^+ \nu_e)$
and $I_K = \Gamma(D^0 \to K^- e^+ \nu_e)\slash\Gamma(D^+ \to \bar{K}^0
e^+ \nu_e)$, with lifetimes input from Ref.~\cite{pdg}. 
Isospin symmetry predicts $I_\pi = 2$ and $I_K = 1$.  

The systematic uncertainty
(see Ref.~\cite{PRD}) is dominated by uncertainty in the number of 
$D\bar{D}$ pairs and in neutrino reconstruction simulation.    
The latter includes inaccuracies in the detector simulation and
in the decay model of the non-signal $D$.
Mainly through use of events with a reconstructed hadronic decay, we evaluate systematic bias, as a function of $q^2$, for
the efficiency of finding and identifying signal hadrons, 
identifying signal electrons, and fake electron rates. 
Similarly, uncertainties that affect the cross-feed rates, such as those
associated with non-signal $\pi^0$ and $\pi^-$
production spectra, as well as $K^-$ faking $\pi^-$, are also assessed as a
function of $q^2$. We correct statistically significant biases, and propagate the uncertainty of each study into our measurement uncertainty.
The remaining systematic uncertainties
include $M_{\mathrm{bc}}$ resolution, the effect
of the single-electron requirement, MC FSR modeling, dependence
on form factors, and the $N_{D\bar{D}}$ determinations. 

Our primary form factor shape analysis utilizes a series
expansion that has been widely advocated as a
physical description of heavy meson form factors~\cite{boyd1,boyd2,grinstein,rhill}:
\begin{equation}
f_+(q^2) = \frac{1}{P(q^2)\phi(q^2,t_0)}\sum_{k=0}^\infty
a_k(t_0)[z(q^2,t_0)]^k. 
\end{equation}
The expansion results from an analytic continuation of the form factor into
the complex $t=q^2$ plane, with a branch cut on the real axis for $t>(M_D+M_{K,\pi})^2$ that is mapped by
$z(t,t_0)=(\sqrt{t_+ - t} - \sqrt{t_+ - t_0})/(\sqrt{t_+ - t} + \sqrt{t_+ - t_0})$ onto the unit circle. The constants $t_\pm \equiv (M_D \pm m_{K,\pi})^2$, and  $t_0$  is the (arbitrary)
$q^2$ value that maps to $z=0$.   The physical region is restricted
to $|z|<1$, so good convergence is expected. 
$P(q^2)$ accommodates sub-threshold resonances:
$P(q^2) = 1$ for $D \to \pi$ and $P(q^2) = z(q^2, M^2_{D^*_s})$ for $D
\to K$.   The function $\phi(q^2,t_0)$ can be any analytic function. We
report $a_k$ parameters that correspond to $t_0 = 0$ and the ``standard''
choice for $\phi$ (see, {\it e.g.} Ref.~\cite{rhill})
that arises naturally in studies of unitarity bounds on $\sum a_k^2$. 

For comparison purposes, we provide results based on the simple and modified
pole models~\cite{BKparam}.  These parameterizations can typically 
accommodate the form factor shapes observed in previous measurements, but
only with parameters that deviate from the underlying physical 
motivation~\cite{hill_fpcp_talk_writeup}.
Note that differing experimental
sensitivities across phase space can result in differing parameter values for a
non-physical parameterization.

\begin{figure}[tb]
\begin{center}
\hspace{0.0cm}
\includegraphics[width=8cm]{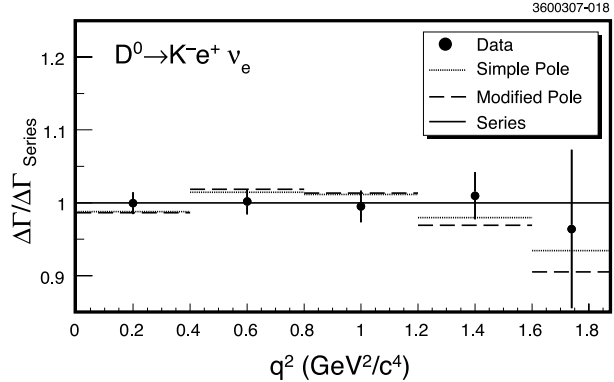} 
\caption{ Fits to the $D^0 \to K^- e^+ \nu_e$ partial branching
 fractions using the three form factor models described in the text. All
 results are normalized to the series expansion result.   }
\label{fig:ff}
\end{center}
\end{figure}

Each parameterization is fit to our measured rates for the five $q^2$ regions; parameter systematic uncertainties are obtained from fits to the
rates obtained for each systematic variation. 
Table~\ref{table:bf} summarizes the results, and  Fig.~\ref{fig:ff}
compares fits for the three parameterizations in our most precise mode
$D^0 \to K^- e^+ \nu_e$.  For the series expansion, we also express our results 
as physical observables: the intercept $|V_{cq}|f_+(0)$ and
$1+1\slash\beta - \delta\propto df_+/dq^2|_{q^2=0}$~\cite{rhill}, which represents the effects of gluon hard-scattering ($\delta$) and scaling
violations ($\beta$).  The results from $D^0$ and $D^+$ decay agree well.

For the series expansion, our kaon data prefer a non-zero quadratic $z$
term. The probability of $\chi^2$ improves from 29\% (22\%) to 89\%
(44\%) with that additional term for the $K^-$ ($\bar{K}^0$)
fit. The pion measurements lack the sensitivity to probe this term,
and two and three parameter fits yield similar results for the first two
parameters.  Since a quadratic term appears preferred for the kaons,
however, we include that term in our series fits to the pion data to
improve the probability that our shape uncertainties bracket the true
form factor shape.  While three of the central values for $a_2$ are an order of
magnitude larger than the other terms, we stress that regions of
parameter space with $a_2$ of similar magnitude to $a_0$ and $a_1$ fall
well within the 90\% hypercontour for the fit, so no conclusion
can be drawn about the size of $a_2$ or (potential lack of) convergence of
the series from these data.

In the simple pole model, we fit for the intercept and the pole mass $m_{\mathrm{pole}}$.
In the modified pole model, we fix the leading pole mass at the physical
value, and fit for the intercept and the parameter $\alpha$, 
which determines the effective higher pole contribution.
We obtain reasonable $\chi^2$ values, but obtain pole masses that deviate
from $M_{D_s^*}$ ($M_{D^*}$) in the kaon (pion) modes by over $3\sigma$ 
for the most precise fits. The $1+1\slash\beta - \delta$ value for the
series expansion fit to 
the $K^- e^+\nu_e$ data  is over $3\sigma$ 
from the value of $\sim$2 necessary for physical
validity of the BK parameterization, while our values
for the BK $\alpha$ parameters from the kaon modes imply $1+1\slash\beta - \delta$ values tens of $\sigma$ away.

We extract $|V_{cd}|$ and $|V_{cs}|$ by combining our $|V_{cq}|f_+(0)$
results from the series expansion fit with the unquenched LQCD 
results \cite{fnalqcd} $f_+^{(D\to\pi)}(0)=0.64(3)(6)$ and 
 $f_+^{(D\to K)}(0)=0.73(3)(7)$ (Table~\ref{table:bf}).  
Averaging the $D^0$ and $D^+$ results (heeding correlations), we find 
$|V_{cd}| = 0.217(9)(4)(23)$ and 
$|V_{cs}| = 1.015(10)(11)(106)$,
with the $f_+$ uncertainty listed last.
The discretization uncertainty in the FNAL LQCD charm quark action dominates.

We also extract the ratio $|V_{cd}|/|V_{cs}|$ from the ratio of our
measured form factors. Averaging over $D^0$ and $D^+$ modes, with
correlations accounted for, gives
$|V_{cd}|f^{(D\to\pi)}(0)\slash |V_{cs}|f^{(D\to K)}(0) = 
0.188(8)(2)$. Comparison to a 
recent light cone sum rules
(LCSR) calculation~\cite{Ball:2006yd} where
$f^{(D\to\pi)}(0)/f^{(D\to K)}(0) = 0.84(4)$, yields
$|V_{cd}|\slash |V_{cs}| = 0.223(10)(3)(11)$.  
 
In summary, we have measured branching fractions and their
ratios for four semileptonic $D$ decay modes in five $q^2$
bins. The branching fraction results are the most precise ever measured and
agree well with world averages.  
Our modified pole $\alpha$ parameter results agree within 1.3$\sigma$
with previous determinations by CLEO III
\cite{cleo_2005}, FOCUS~\cite{focus_2005}, and $Ke\nu$
results from Belle~\cite{belle_2006}, but show over $3\sigma$ disagreement
with Belle $K\mu\nu$ results and LQCD fits.  The $\alpha$ parameters
obtained with our individual $Ke\nu$ results are separated from the
recent BaBar result \cite{Aubert:2006mc} by about $2.5\sigma$.  
Our $z$ expansion results agree with BaBar's at about the $2\sigma$ level, 
depending on the total level of correlation between the BaBar $r_1$ and $r_2$ parameters. We have made the most precise CKM
determinations from $D$ semileptonic decays to date, and the results
agree well with neutrino-based determinations of $|V_{cd}|$ and
charm-tagged $W$ decay measurements of $|V_{cs}|$~\cite{pdg}.  Overall, 
these measurements represent a marked improvement in our knowledge
of $D$ semileptonic decay.

We gratefully acknowledge the effort of the CESR staff 
in providing us with excellent luminosity and running conditions. 
This work was supported by 
the A.P.~Sloan Foundation,
the National Science Foundation,
the U.S. Department of Energy, and
the Natural Sciences and Engineering Research Council of Canada.

\end{document}